# New evidence for structural and magnetic properties of GaAs:(Mn,Ga)As granular layers

J. Bak-Misiuk[a], K. Lawniczak-Jablonska[a], E. Dynowska[a], P. Romanowski[a],
J. Z. Domagala[a], J. Libera[a], A. Wolska[a], M. T. Klepka[a], P. Dluzewski[a],
J. Sadowski[a,b], A. Barcz[a], D. Wasik[c], A. Twardowski[c], A. Kwiatkowski[c], W. Caliebe[d]

[a]Institute of Physics PAS, al. Lotnikow 32/46, PL-02668 Warsaw, Poland
[b]MAX-Lab, Lund University P.O. Box. 118, S-22100 Lund, Sweden
[c]Institute of Experimental Physics, Faculty of Physics, Univ. of Warsaw, ul. Hoza 69, PL-00681 Warsaw, Poland
[d]HASYLAB at DESY, Notkestr. 85, D-22603 Hamburg, Germany

**Abstract**

Structural and magnetic properties of GaAs thin films with embedded MnAs nanoclusters were investigated as function of the annealing temperature and layers composition. The presence of two kinds of nanoclusters with different dimensions and structure were detected. The fraction of Mn atoms in each kind of cluster was estimated from the extended X-ray absorption fine structure analysis. This analysis ruled out the possibility of the existence of nanoclusters containing a hypothetic MnAs cubic compound - only (Mn,Ga)As cubic clusters were detected. Change of the layer strain from the compressive to tensile was related to the fraction of zinc blende and hexagonal inclusions. Thus the zinc blende inclusions introduce much larger strain than hexagonal ones. The explanation of observed thermal induced strain changes of the layers from the compressive to tensile is proposed. The magnetic properties of the samples were consistent with structural study results. Their showed that in sample containing solely cubic (Mn,Ga)As inclusions Mn ions inside the inclusions are still ferromagnetically coupled, even at room temperature. This fact can be explained by existence in these clusters of GaMnAs solid solution with content of Mn higher than 15 % as was found in theoretical calculations.

## 1. Introduction

Magnetic semiconductors with carrier-induced ferromagnetism have received a lot of interest, since they hold out prospects for using electron spins in electronic devices. $Ga_{1-x}Mn_xAs$ is one of the diluted magnetic semiconductor materials (DMS) which has been the most extensively studied after its discovery in 1996 by Ohno et. al. [1]. It has attracted a great deal of attention both as an object of basic studies of DMS systems [2-4] and as a versatile material suitable for testing prototype spintronic devices [5-8]. The low solubility limit of Mn in GaAs (less than 0.1%) has been overcome by use of a low temperature molecular beam epitaxial (MBE) growth, for preparation of monocrystalline GaMnAs layers. MBE is a highly nonequilibrium crystal growth method; moreover the use of extremely low (as for GaAs) growth temperatures, in the range of 170-250 °C, allows the creation of GaMnAs ternary alloy with Mn content up to 20% [9,10]. Mn located at Ga sites in GaAs host is an acceptor and gives a very large concentration of holes (in the range of $10^{20}$ to $10^{21}$ cm$^{-3}$) in GaMnAs with Mn content in the range of a few percent. However, due to the low crystallization temperature, GaMnAs contains point defects retaining donor character and partially compensating the $Mn_{Ga}$ acceptors. The most important defects of this kind are As antisites ($As_{Ga}$) and Mn interstitials ($Mn_I$). In order to optimize the properties of GaMnAs such as conductivity and ferromagnetic phase transition temperature ($T_c$) it is necessary to minimize the concentration of these two compensating defects. The concentration of $As_{Ga}$ can be minimized by a choice of suitable conditions for low temperature MBE growth [10]. On the other hand, the concentration of $Mn_I$ defects can be highly reduced by the use of low temperature (170 - 250 °C) post-growth annealing procedures [11-13].

Low temperature (*LT*) annealing of thin GaMnAs epitaxial layers at temperatures up to 250 °C causes a decrease of the layer lattice parameter due to removal of the Mn interstials which form double donors [14]. A reduction of Mn interstitials leads to enhanced carrier concentration and Curie temperature. The present $T_c$ record in the diluted $Ga_{1-x}Mn_xAs$ material is about 190 K [15], which is remarkably high for a DMS material but still too low for application purposes.

On the other hand, it has been demonstrated that, as an effect of high temperatures *(HT)* annealing at 400 – 700 °C of the GaMnAs layers [16-21] or the Mn-implanted GaAs crystals [22] ferromagnetic MnAs precipitates are fairly easily produced, yielding multi-phase materials. The granular GaAs:(Mn,Ga)As material exhibits a ferromagnetic/superparamagnetic behavior at room temperature, depending on the cluster size. However, in the case of composite systems containing ferromagnetic inclusions (Mn,Ga)As in the semiconductor crystal matrix (GaAs), the main advantage of a DMS system - the ability to control magnetic properties by such factors as flow of current, irradiation with light, and/or application of an external electric field-is lost. On the other hand, such composite systems are currently thoroughly investigated in search of possibilities of overcoming these drawbacks [23-27]. In this context we believe that it is still interesting to study the properties of the GaAs:(Mn,Ga)As system.

Depending on the annealing conditions, small cubic zinc-blende type (ZB) or larger NiAs-type MnAs nanoclusters are created due to HT annealing of GaMnAs [16, 20, 21]. It has been found [16, 20, 28] that creation of granular GaAs:(Mn,Ga)As is accompanied by a contraction of the layer lattice parameters; the out-of-plane lattice constant is usually smaller than that of the GaAs substrate.

Strong ferromagnetic behavior was achieved only when hexagonal clusters are formed in the GaAs matrix [16, 20, 21]. The magnetic, optical and electron transport properties of these composite materials depend strongly on the microscopic structure of the precipitates. Therefore, it is very important to find correlations between annealing conditions, Mn concentration, and the precipitates structure.

The aim of this work was to study systematically the influence of post–annealing conditions and Mn contents on the formation of nanoclusters and related values of strains in GaAs:(Mn,Ga)As layers. The X-ray diffraction (XRD) and transmission electron microscopy (TEM) methods were applied to get full information about the structure and dimensions distribution of clusters. Neither of these methods provides knowledge about the chemical composition of clusters or the amount of Mn being bonded in each of the detected kind of nanoclusters. Therefore, X-ray absorption spectroscopy (XAS) was also applied. To control the Mn content and depth profile after annealing, secondary ion mass spectroscopy (SIMS) was used. Finally, the magnetic properties of samples were measured and related to structural properties.

**2. Experimental**

$Ga_{1-x}Mn_xAs$ layers with nominal Mn contents $x = 0.015, 0.06, 0.07$, and $0.08$ were grown by MBE on the (001)-oriented GaAs substrates at a temperature of 230 °C. The nominal thickness of the layers was 0.8 μm. After deposition each sample was cut into three pieces: one of them was left untreated, the others were reintroduced into the MBE growth chamber and annealed under $As_2$ flux for 30 min at $T = 500$ °C and at $T = 600$ °C, respectively.

To estimate the real Mn contents in $Ga_{1-x}Mn_xAs$ layers and to check the influence of annealing on the distribution of Mn atoms inside them, SIMS measurements by use the CAMECA IMS6F micro-analyzer were carried out. These measurements were performed with an oxygen ($O_2^+$) primary beam, with the current kept at 600 nA. The size of the eroded crater was about 150 μm × 150 μm and the secondary ions were collected from the central region of 60 μm in diameter. The Mn content was derived from the intensity of the $Mn^+$ species and the matrix signal, $As^+$, was taken as a reference. Mn implanted GaAs was used as a calibration standard. The accuracy of the Mn content determination in the SIMS method is ~5 %. The thickness of the examined layers was measured by an Alpha-Step Profiler.

The X-ray measurements were performed by using a high resolution Philips material research diffractometer (MRD) in double and triple crystal configuration, equipped with a standard laboratory source of Cu K$\alpha_1$ radiation. Moreover, the measurements with the use of monochromatic synchrotron radiation ($\lambda = 1.54056$ Å) at the W1 beamline at DESY-Hasylab were also performed.

The out-of-plane ($a_\perp$) and in-plane ($a_\parallel$) lattice parameters were calculated from $2\theta/\omega$ in the vicinity of the 004 symmetrical and 224 asymmetrical reflections. The relaxed lattice parameter ($a_{rel}$) as well as the in-plane strain ($\varepsilon$) for each layer were calculated according to the formulas:

$$a_{rel} = \frac{a_\perp + \nu a_\parallel}{1+\nu}, \quad (1)$$

where $\nu = 2\dfrac{c_{12}}{c_{11}}$ - Poisson's ratio for cubic crystals;

$$\varepsilon = \frac{a_\parallel - a_{rel}}{a_{rel}} \quad (2)$$

The determination of the lattice parameters of these MnAs nanoclusters was possible using synchrotron radiation. The orientation of the MnAs nanoclusters embedded in GaAs matrix is well–defined [17]: (00.1) planes of MnAs are parallel to {111} GaAs planes, the (01.0) MnAs directions are parallel to <1 $\bar{1}$2> GaAs directions, and the <11.0> MnAs directions are parallel to <1 $\bar{1}$0> or <11.0> GaAs directions. Therefore, the lattice parameters of MnAs were calculated from 02.2 and 03.0 reflections which were detected by the $\omega$ scan in the vicinity of symmetrical 004 GaAs reflection, and from the $2\theta/\omega$ scan around the asymmetrical 224 GaAs reflection.

TEM studies were carried out using a JEOL 2000EX instrument operating at 200 kV accelerating voltage. The cross-sectional specimens were mechanically polished, dimpled and finally milled with $Ar^+$ ions beams at 2 kV and 8° inclined to the sample surface. The histograms of nanocluster sizes were obtained from high-resolution images obtained in the <011> projection.

XAS measurements were performed at liquid nitrogen temperature, at the A1 experimental station in DESY-HASYLAB using a double crystal Si (111) monochromator. The Mn K-edge spectra were registered using a seven-element fluorescence Si detector. To reduce the error introduced by the diffraction peaks, which are present in the well oriented crystalline samples for each sample several spectra were collected at slightly different angles around 45°. The spectra were averaged and then analyzed.

Magnetization measurements were performed using superconducting quantum interference device (SQUID) magnetometer as a function of temperature (5-325 K) at constant magnetic fields. Magnetic field was applied in-plane of the sample surface. Magnetization data were corrected for the diamagnetic contribution of GaAs substrate. More details of the experiment will be published elsewhere.

## 3. Results and Discussion

*3.1 Secondary ion mass spectroscopy studies*

The concentration of Mn and the thickness of the $Ga_{1-x}Mn_xAs$ layers found from SIMS measurements in the as-grown samples and the relation of these concentrations to the nominal Mn concentrations are presented in the Fig. 1.

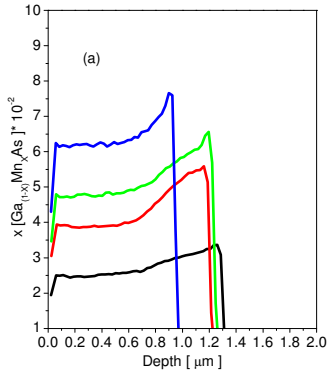

Fig. 1. (Color online) The Mn concentration in-depth profiles and layer thickness derived from SIMS study of the investigated samples.

The concentration of Mn and thickness of the grown layers differ from nominal values. The smallest content of Mn was $x = 0.025$ and the highest $x = 0.063$. Therefore, the relation of nominal contents $x = 0.015$, 0.06, 0.07 and 0.08 to SIMS values is $x = 0.025$, 0.04, 0.05 and 0.063 respectively. In the further discussion we will use the concentration values as estimated by SIMS. The concentration of Mn in all of the examined samples did not change significantly during the annealing but some tendency of the Mn accumulation at the interface with the GaAs substrate was observed. Results of the SIMS measurements for the annealed sample $x = 0.04$ are shown in Fig. 2 as an example. During the annealing at 500 °C and 600 °C Mn atoms did not diffuse significantly to the surface or to the substrate.

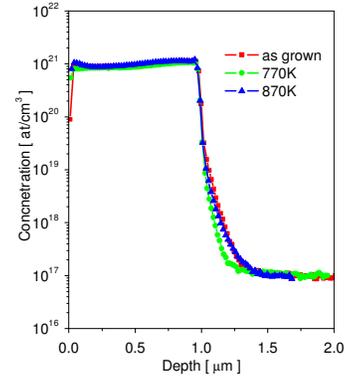

Fig. 2. (Color on line) The concentration of Mn atoms in the sample with $x = 0.04$ as-grown and after annealing.

*3.2 X-ray diffraction studies*

The lattice parameters of the investigated layers were calculated from the symmetrical and asymmetrical $2\theta/\omega$ scans performed by the high resolution MRD diffractometer. It was confirmed that all $Ga_{1-x}Mn_xAs$ layers grow on the GaAs substrate pseudomorphically, i.e. the in-plane lattice parameter ($a_{||}$) is equal to that of the GaAs substrate. It was found that the annealing of these layers does not change their in-plane lattice parameter.

The $2\theta/\omega$ scans around 004 GaAs reflection for the as-grown and annealed samples with low ($x = 0.025$) and higher Mn concentration ($x = 0.05$) are presented in Fig. 3 and Fig. 4, respectively. The 004 peaks marked by 1, 2, and 3 originate from layers as-grown, annealed at 500 °C, and annealed at 600 °C, respectively.

The position of the 004 peak from the as-grown layer is situated on the left side of the 004 peak from the GaAs substrate which means that this layer is compressively strained ($a_{\perp layers} > a_{GaAs}$). The relatively large compressive strain of the as-grown layers results from Mn interstitial and $As_{Ga}$ anti-site defects as well as from Mn in Ga sites [29, 30]. It was also found that for the as-grown layers an increase of Mn content leads to an increase of the $a_\perp$ lattice parameter, and, in consequence, to an increase of the strain state ($\varepsilon$). The relaxed lattice parameters ($a_{rel}$) of these layers are larger than the lattice parameter of GaAs substrate. On the other hand, the positions of 004 peaks originating from the annealed layers are shifted towards higher angles with respect to the position of 004 GaAs peak, which indicates the change of strain state from compressive to tensile ($a_{\perp layers} < a_{GaAs}$). Simultaneously, the relaxed lattice parameters of these layers become smaller than the lattice parameter of GaAs. In the case of lower Mn content ($Ga_{0.975}Mn_{0.025}As$) the annealing at higher temperature leads to further decrease of the $a_\perp$ lattice parameter (Fig. 3) which results in further decrease of $a_{rel}$ and increase of tensile strain. The opposite effect was found for the layers with higher Mn concentration ($Ga_{0.95}Mn_{0.05}As$) - an increase of annealing temperature to 600 °C leads to an increase of the $a_\perp$ lattice parameter and a decrease of tensile strain. The calculated values of the lattice parameters $a_\perp$ and $a_{rel}$, and the in-plane strain state ($\varepsilon$) of the layers are given in Table 1.

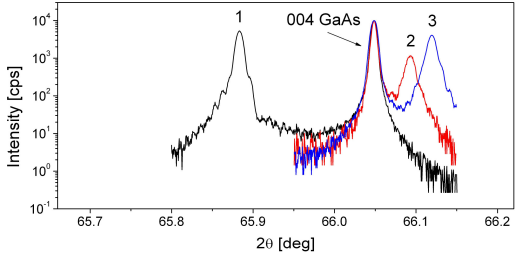

Fig. 3. (Color on line) The $2\theta/\omega$ scans of 004 reflection for $Ga_{0.975}Mn_{0.025}As$: (1) – as-grown, (2) – after annealing at 500°C, (3) – after annealing at 600 °C.

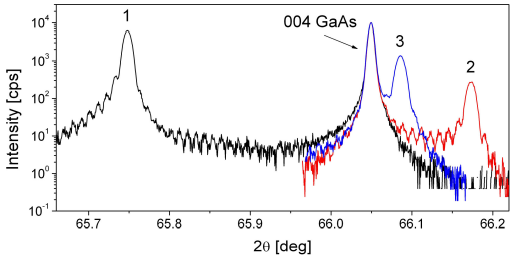

Fig. 4. (Color on line) The $2\theta/\omega$ scans of 004 reflection for $Ga_{0.95}Mn_{0.05}As$ : (1) – as-grown, (2) – after annealing at 500°C, (3) – after annealing at 600 °C.

The decrease of the lattice parameter of annealed GaMnAs layers is related to the formation of granular GaAs:(Mn,Ga)As material consisting of MnAs nanoclusters [16, 20, 28]. The mechanism of such behavior is not clear up to now, and therefore we made an attempt to explain this effect. In our studies the evidence of such nanoclusters was possible owing to the use of synchrotron radiation. The $\omega$ scan recorded around the 004 GaAs reflection (Fig. 5a) allowed the detection of two reflections of 02.2 originating from hexagonal, NiAs-type MnAs nanoclusters (Bragg angles for these reflections are very similar: $\theta = 33.30°$ for 02.2 of MnAs bulk and $\theta = 33.03°$ for 004 GaAs). The broadening of the 02.2 peaks can be attributed to small degrees of disorder in the orientations of different MnAs nanoclusters. From the $2\theta/\omega$ scans of 02.2 MnAs reflection (Fig. 5a, inset) and the asymmetrical reflection of $22\bar{4}$ GaAs (Fig. 5b), the lattice spacings $d_{02.2} = 1.406$ Å and $d_{03.0} = 1.071$ Å were calculated. On the basis of these values the lattice parameters of hexagonal unit cell of MnAs nanoclusters, $a = 3.710 \pm 0.002$ Å and $c = 5.812 \pm 0.007$Å, were determined. These values are closed to those reported in (Ref.18).

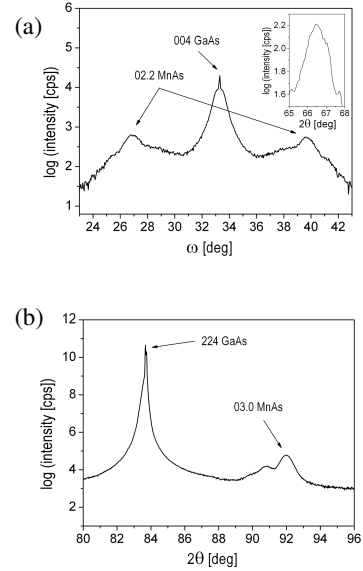

Fig. 5. The X-ray patterns for $Ga_{0.95}Mn_{0.05}As$ after annealing at 600 °C. The $\omega$ scan around 004 GaAs reflection – (a); the inset displays the $2\theta/\omega$ scan of 02.2 reflection and (b) – $2\theta/\omega$ scan in the vicinity of 224 GaAs reflection.

The crystal structure of the bulk MnAs is hexagonal of NiAs-type with lattice parameters $a = 3.7187$ Å and $c = 5.7024$ Å (e.g., Ref. 31). Differences between these values and those determined for MnAs nanoclusters results from both the lattice mismatch between the lattice parameters of MnAs and GaAs and the large difference between the thermal expansion coefficients for these compounds. Due to these effects, the oriented hexagonal MnAs clusters are distorted with respect to unstrained bulk MnAs. This distortion creates the biaxial strain ($\varepsilon_{11} = \varepsilon_{22}$) in {00.1} planes and in along the c-axis ($\varepsilon_{33}$) in <00.1> directions. The values of these strains are $\varepsilon_{11} = \varepsilon_{22} = -0.2 \cdot 10^{-2}$, and $\varepsilon_{33} = 1.8 \cdot 10^{-2}$. This means that the $a$ lattice parameter of nanoclusters is compressed by only 0.20%, while the $c$ lattice parameter is increased by 1.8 %. Knowing the strains of the nanoclusters we can estimate the related stresses, $\delta_a$ and $\delta_c$, through Hooke's law using the elastic constants for hexagonal MnAs [32]. The stresses $\delta_a$ and $\delta_c$, along the $a$ and $c$ axes, respectively, are $\delta_a = -0.04$ GPa and $\delta_c = 1.92$ GPa. The same stresses, $\delta_a$ and $\delta_c$, but with opposite sign, act in the <110> and <111> directions of the matrix unit cell, decreasing its lattice parameter. The stress $\delta_a$ is very small compared with that acting in the <111> directions, $\delta_c$; therefore in further considerations we will take into account only $\delta_c = -1.92$ GPa. From this value of stress, using the bulk modulus (76 GPa) for GaAs and pressure $p = 1.92$ GPa, the lattice parameter of the hydrostatically compressed GaAs unit cell $a_{strained} = 5.6057$Å was determined. However, this value is much smaller than the experimental values, $a_{rel}$, obtained from our measurements for all investigated samples (see Table 1). Because the X-ray measurements give an average value of the lattice parameters of strained and unstrained unit cells, we can conclude that only part of the matrix unit cells is strained by hexagonal MnAs inclusions. The fraction of the strained unit cells (y) can be estimated from the simple equation

$$y \times a_{GaAs\ strained} + (1-y) \times a_{GaAs} = a_{average} \qquad (3)$$

where $a_{average}$ is the relaxed lattice parameter ($a_{rel}$) obtained from measurement of the $Ga_{1-x}Mn_xAs$ layer after annealing, $a_{GaAs}$ is the lattice parameter of unstrained GaAs unit cells.

Table 1. Values of the lattice parameters $a_\perp$, $a_{rel}$ and in-plane strain $\varepsilon$, for $Ga_{(1-x)}Mn_xAs$ layers before and after annealing. The in-plane lattice parameter $a_\parallel$ for all studied layers is the same as that of GaAs substrate: $a_\parallel = a_{GaAs} = 5.6533$ Å. All lattice parameters were calculated with an accuracy of $10^{-4}$ Å.

| Sample $Ga_{1-x}Mn_xAs$ | $a_{\perp\text{as-grown}}$ [Å] | $a_{\perp\,500°C}$ [Å] | $a_{\perp\,600°C}$ [Å] | $a_{relax}$ as-grown | $a_{relax}$ 500 °C | $a_{relax}$ 600 °C | $\varepsilon_{\text{as-grown}} \times 10^{-4}$ | $\varepsilon_{500°C} \times 10^{-4}$ | $\varepsilon_{600°C} \times 10^{-4}$ |
|---|---|---|---|---|---|---|---|---|---|
| $x = 0.025$ | 5.6658 | 5.6500 | 5.6479 | 5.6596 | 5.6517 | 5.6506 | -11.1 | 2.8 | 4.8 |
| $x = 0.04$ | 5.6732 | 5.6471 | 5.6493 | 5.6633 | 5.6502 | 5.6513 | -17.7 | 5.5 | 3.5 |
| $x = 0.05$ | 5.6763 | 5.6439 | 5.6505 | 5.6648 | 5.6486 | 5.6519 | -20.3 | 8.3 | 2.5 |
| $x = 0.063$ | 5.6857 | 5.6475 | 5.6501 | 5.6695 | 5.6504 | 5.6517 | -28.6 | 5.13 | 2.8 |

In particular, for the $Ga_{0.95}Mn_{0.05}As$ layer annealed at 600 °C, ($a_{rel} = 5.6519$ Å) the fraction of the strained unit cells inside the GaAs matrix is $y = 0.03$. This estimation was performed assuming the existence of only hexagonal MnAs inclusions, although according to other studies (TEM, EXAFS) nanoparticles of cubic (Mn,Ga)As are also created. The lattice parameter reported for cubic MnAs is ~ 5.9 Å (e.g., Ref. 28). Unfortunately, the size of these inclusions is too small for detection by X-ray diffraction. As was shown in TEM studies, the creation of the inclusions starts from the metastable, cubic small nanoclusters, which grow epitaxially on {111} GaAs planes. Calculation of the stress developed in the GaAs matrix by cubic (Mn,Ga)As nanoinclusions, using the above described method for GaAs:MnAs consisting of hexagonal MnAs is impossible, because neither the strain state nor the elastic constant of cubic (Mn,Ga)As is known.

3.3 *Transmission electron microscopy studies*

For all the annealed samples, nanoclusters were observed. Fig. 6 shows cross-sectional TEM images of granular layers for samples $Ga_{0.975}Mn_{0.025}As$ and $Ga_{0.937}Mn_{0.063}As$ annealed at 600 °C. It was found that small clusters with sizes less than 10 nm have cubic ZB structure in contrast to the large nanoclusters, which have NiAs-type hexagonal structure. In both cases nanoclusters were coherently aligned with the GaAs matrix (Fig. 6c and 6d).

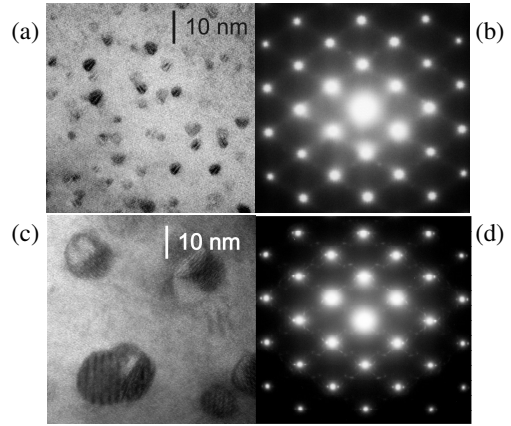

Fig. 6. Cross-sectional bright-field TEM images and electron diffraction patterns of granular GaAs:(Mn,Ga)As layers obtained after annealing of $Ga_{1-x}Mn_xAs$ with two different Mn compositions $x$ at 600 °C: (a), (b) – $x = 0.025$; and (c), (d) – $x = 0.063$.

The Bragg spots from GaAs structure were visible on all the difraction patterns. For $Ga_{0.975}Mn_{0.025}As$ (Fig. 6c) lines of weak diffuse intensities along <111> directions were present. These originated from nucleation of (Mn,Ga)As cubic clusters on of the {111} planes of the GaAs matrix. On the diffraction pattern for the $Ga_{0.937}Mn_{0.063}As$ sample (Fig. 6d) in addition to the spots from GaAs matrix week spots attributed to hexagonal MnAs clusters were visible. From analysis of the electron diffraction patterns it was found that the additional spots originated from NiAs-type hexagonal nanocrystals with {00.1} lattice planes parallel to the {111} of GaAs matrix and with MnAs <11.0> directions collinear with <110> of GaAs.

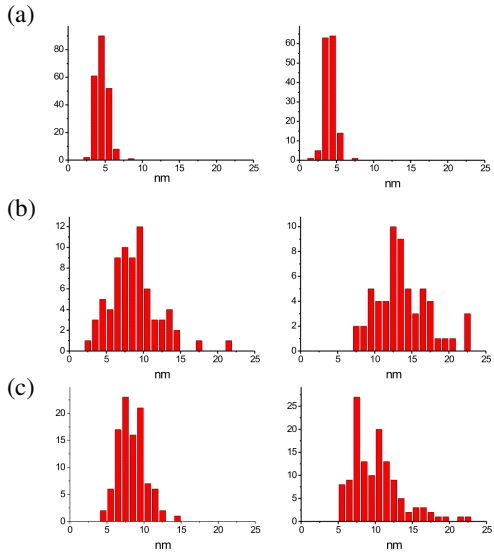

Fig. 7. Histograms of $Ga_{0.975}$ $Mn_{0.025}As$ (a), $Ga_{0.95}$ $Mn_{0.05}As$ (b), and $Ga_{0.937}$ $Mn_{0.063}As$ (c) annealed at 500 °C (left column) and 600 °C (right column).

The distribution of nanocluster diameters is presented in Fig. 7. For the sample with the lowest Mn concentration ($x = 0.025$), after annealing at 500 °C and 600 °C only small nanoclusters with diameter about 3-6 nm were found. For the sample $Ga_{0.95}$ $Mn_{0.05}As$ most of the nanoclusters have diameter between 6 and 11 nm already after annealing at 500 °C. After annealing at 600 °C only a few nanoclusters with diameter smaller than 10 are left; most of the nanoclusters have larger diameters. In the case of the $Ga_{0.937}$ $Mn_{0.063}As$ sample, after annealing at 500 °C, only a small number of nanoclusters with diameter larger than 10 nm was found, and even after annealing at 600 °C a relatively large number of nanoclusters with small diameter were still detected.

The X-ray diffraction results were compared with those obtained by TEM measurements. For the sample with the lowest Mn concentration, $Ga_{0.975}Mn_{0.025}As$, no precipitations were detected by X-rays, but the TEM image (Fig. 7) clearly shows the cubic nanoculsters with dominating diameters of about 3-6 nm. Simultaneously, after annealing at 600 °C the increase of the layer strain was detected by X-ray measurements (Table 1). An increase of the strain state of the layer after annealing at 600 °C is related to the further ordering of Mn in the cubic clusters. The opposite effect was observed for the layers with larger Mn concentrations ($x = 0.05$, and 0.063): here annealing at higher temperature leads to creation of hexagonal MnAs clusters (see Figs 6 and 7) which results in a decrease of the strain state of the layers (Table 1). However, the observed decrease of the layer strain was less pronounced for the $Ga_{0.937}Mn_{0.063}As$ sample, which is connected with relatively larger number of cubic nanoclusters in this sample with respect to the $Ga_{0.95}Mn_{0.05}As$ one (see Fig. 7).

The careful analysis of X-ray and TEM results performed in our work leads to the conclusion that the cubic inclusions act with larger stress on the matrix unit cells than the hexagonal ones. However, not only the structure of the clusters but also the content of Mn in the cubic nanoinclusions can be responsible for the strain changes of the layers.

*3.4 X-ray absorption studies*

The XRD and TEM measurements provide information about the crystallographic structure of the matrix and formed nanoclusters but do not allow to estimate the part of Mn atoms located in each kind of nanoclusters or their chemical composition. XAS probes the local atomic order around the absorbing atom; therefore, it is an ideal tool for the determination of the Mn location in the investigated structures [28, 34, 35, 36]. The EXAFS study of the GaMnAs samples after *HT* annealing was performed in order to estimate the fraction of Mn atoms located in the ZB and hexagonal nanoclusters. The XAS spectra were analyzed using the IFEFFIT package [37]. The used mathematical formalism represents a Real-Space Multiple-Scattering (RSMS) with summation of contributions from all single scattering paths of inner photoelectrons emitted during the absorption process.

In the analysis of the EXAFS spectra Mn based nanoclusters with ZB- and NiAs-type hexagonal structure immersed in the GaAs matrix were considered. Different values of lattice constants for hypothetical zinc-blende MnAs have been reported in the literature, namely 5.7 Å, 5.9 Å and 5.98 Å [38-41]. Corresponding Mn-As bond lengths are ~2.47 Å, ~2.55 Å and ~2.59 Å. Therefore, as a starting model it was assumed that the distance of the first coordination sphere is equal to 2.55 Å. For the hexagonal inclusions, the NiAs-type structure (P6$_3$/mmc space group) with the lattice constants $a = 3.7187$ Å and $c = 5.7024$ Å [31] was considered.

The amplitude-damping factor ($S_0^2$) was estimated similarly as in (Ref. 28) and found to be 0.85, that value was used in all fittings. The data were analyzed from $k_{min}$ around 2 Å$^{-1}$ to $k_{max}$ from 12.8 to 14.5 Å$^{-1}$ depending on data quality. The fitting was performed in R space. The data were weighted by $k^3$ to enhance the oscillations for high k.

To keep the number of fitting parameters as small as possible, the parameter weighting the number of atoms in the coordination spheres according to the fraction of Mn atoms in each kind of nanoclusters was introduced. N1 represents the fraction (in %) of the nanoclusters with the cubic structure. The fraction of the nanoclusters with the hexagonal structure was assumed to be 100 - N1.

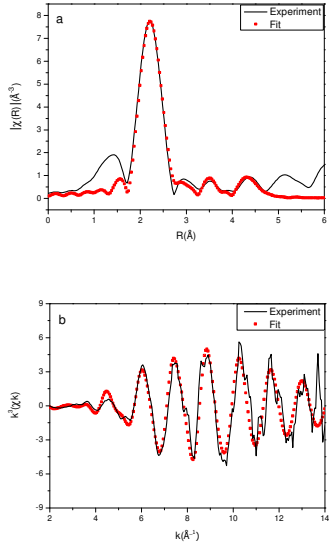

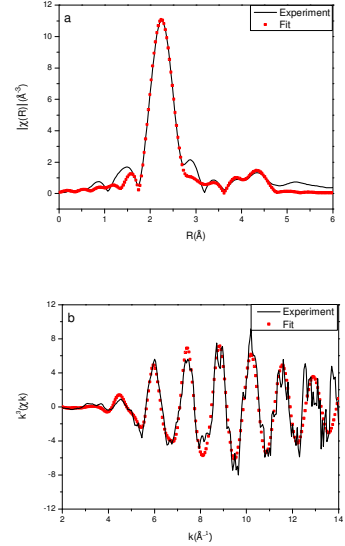

Fig. 8. $Ga_{0.975}Mn_{0.025}As$ annealed at 500 °C. Model of (Mn,Ga)As ZB inclusions. (a) Fourier Transform of the EXAFS (squares) and the result of fitting (full line). (b) EXAFS oscillations (squares) and the result of fitting (full line).

Taking into consideration the existence of hypothetical zinc blende MnAs, widely discussed in the literature, we have constructed a model of such a structure with the lattice parameter of 5.9 Å. According to the atomic order in this structure, Mn atoms should have 4 As atoms as the nearest neighbors and, in the next shells, 12 Mn and 12 As atoms. Surprisingly, for none of the investigated samples the Mn atoms can be found in the second coordination sphere. Only Ga atoms in the second sphere provided physical parameters of the fit. We have also considered the possibility of the coexistence (Mn,Ga)As solid solution and cubic MnAs nanoclusters. In all samples the fraction of MnAs ZB was at the level of error. Therefore, we can rule out the existence of hypothetic pure MnAs nanoclusters in the investigated samples. Only nanoclusters of (Mn,Ga)As zinc blende solid solution were found. The atomic distances and disorder parameters differ in these nanoclusters and depend on the Mn content and annealing temperature (see Table 2 and 3). This can explain the variety of lattice parameter values reported in the literature for hypothetic ZB MnAs and can be related to the dimension of clusters and the stress induced by the matrix.

In the case of the hexagonal structure, the distances between the central atom and specific shells were kept according to crystallographic data. The sigma square parameters were fitted for all samples taking into account the fact that chemical disorder can differ from sample to sample. The numerical results of such fitting are collected in Table 2 for samples annealed at 500 °C and in Table 3 for samples annealed at 600 °C. The fractions of Mn atoms found in each phase are summarized in Table 4. Some of the experimental data together with fitted models for the investigated samples are shown in Figures 8-13.

Fig. 9. $Ga_{0.975}Mn_{0.025}As$ annealed at 600 °C. Model of (Mn,Ga)As ZB inclusions. (a) Fourier Transform of the EXAFS (squares) and the result of fitting (full line). (b) EXAFS oscillations (squares) and the result of fitting (full line).

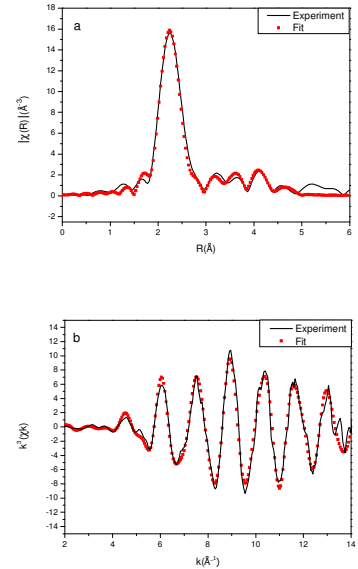

Fig. 10. $Ga_{0.95}Mn_{0.05}As$ annealed at 500 °C. Model of 53 % (Mn,Ga)As ZB and 47% MnAs hexagonal inclusions. (a) Fourier Transform of the EXAFS (squares) and the result of fitting (full line). (b) EXAFS oscillations (squares) and the result of fitting (full line).

Table 2. The inter-atomic distances R in the subsequent coordination spheres, and disorder parameters Sigma as estimated for samples after annealing at 500 °C. Rf denotes the parameter describing the quality of fitting, N1 the fraction of nanocluster in zinc blende structure.

| Atom | R | Sigma | R | Sigma | R | Sigma | R | Sigma |
|---|---|---|---|---|---|---|---|---|
| Sample | $x = 0.025$ | | $x = 0.04$ | | $x = 0.05$ | | $x = 0.063$ | |
|  | Rf=0.0005 | | Rf=0.0001 | | Rf=0.0003 | | Rf=0.0002 | |
| N1 | 100% | | 72(2)% | | 53(5)% | | 69 (2)% | |
| As 4 | 2.557(2) | 0.0055(6) | 2.52(1) | 0.008(1) | 2.55(1) | 0.005(1) | 2.53(1) | 0.006(1) |
| Ga 12 | 4.06(4) | 0.025(6) | 4.15(7) | 0.034(8) | 4.07(4) | 0.013(5) | 4.10(4) | 0.024(5) |
| As 12 | 4.73(3) | 0.018(4) | 4.49(6) | 0.029(7) | 4.27(5) | 0.011(4) | 4.48(3) | 0.018(3) |
| 1-N1 | 0% | | 28% | | 47% | | 31% | |
| As 6 | | | 2.58 | 0.003(2) | 2.58 | 0.006(1) | 2.58 | 0.002(1) |
| Mn 2 | | | 2.85 | 0.02(1) | 2.85 | 0.005(3) | 2.85 | 0.005(4) |
| Mn 6 | | | 3.72 | 0.015(8) | 3.72 | 0.013(9) | 3.72 | 0.006(5) |
| As 6 | | | 4.52 | 0.013(9) | 4.52 | 0.011(6) | 4.52 | 0.02(1) |
| Mn 12 | | | 4.69 | 0.002(1) | 4.69 | 0.011(9) | 4.69 | 0.004(1) |
| As 6 | | | 4.79 | 0.0012(8) | 4.79 | 0.007(4) | 4.79 | 0.003(1) |

Table 3. The inter-atomic distances R in the subsequent coordination spheres and the disorder parameters Sigma as estimated for samples after annealing at 600 °C. Rf denotes the parameter describing the quality of fitting, N1 the fraction of nanocluster in zinc blende structure.

| Atom | R | Sigma | R | Sigma | R | Sigma | R | Sigma |
|---|---|---|---|---|---|---|---|---|
| Sample | $x = 0.025$ | | $x = 0.04$ | | $x = 0.05$ | | $x = 0.063$ | |
|  | Rf =0.0002 | | Rf=0.0001 | | Rf=0.0002 | | Rf=0.0003 | |
| N1 | 100% | | 40(5)% | | 15(5)% | | 50(5)% | |
| As 4 | 2.577(4) | 0.0049(2) | 2.54(1) | 0.006(1) | 2.55 | 0.002(1) | 2.57(1) | 0.006(1) |
| Ga 12 | 4.23(4) | 0.025(7) | 4.13(4) | 0.012(6) | 4.10 | 0.006(4) | 4.19(3) | 0.017(4) |
| As 12 | 4. 68(2) | 0.017(3) | 4.29(7) | 0.014(8) | 4.24 | 0.005(3) | 4.73(5) | 0.02(1) |
| 1-N1 | 0 % | | 60% | | 85% | | 50% | |
| As 6 | | | 2.58 | 0.005(1) | 2.58 | 0.0047(2) | 2.58 | 0.004(1) |
| Mn 2 | | | 2.85 | 0.010(4) | 2.85 | 0.007(3) | 2.85 | 0.005(2) |
| Mn 6 | | | 3.72 | 0.017(5) | 3.72 | 0.011(3) | 3.72 | 0.010(3) |
| As 6 | | | 4.52 | 0.026(9) | 4.52 | 0.012(4) | 4.52 | 0.026(9) |
| Mn 12 | | | 4.69 | 0.0010(3) | 4.69 | 0.009(3) | 4.69 | 0.009(3) |
| As 6 | | | 4.79 | 0.006(3) | 4.79 | 0.006(2) | 4.79 | 0.005(3) |

EXAFS analysis of the samples with different Mn content and after different annealing procedures excluded the existence of MnAs ZB nanoclusters in GaAs matrix. In all of the investigated samples the ZB (Mn,Ga)As nanoclusters were found. Due to small dimension of detected ZB nanoclusters (up to 6 nm) the Ga atoms in second sphere were detected in all samples. The ratio of unit cells exposed to the surface to those in bulk is about 0.5 for clusters up to 5 nm; therefore half of the Mn atoms would see Ga from the surface. However, if MnAs ZB is formed inside the cluster it would be visible in EXAFS as well. Nevertheless, when the dimensions of clusters reach 10 nm and the influence of the surface is negligible, hexagonal inclusions are formed. In the sample with the smallest Mn concentration ($x = 0.025$) annealed at 500 and 600 °C, only ZB phase inclusions were found in agreement with TEM observation.

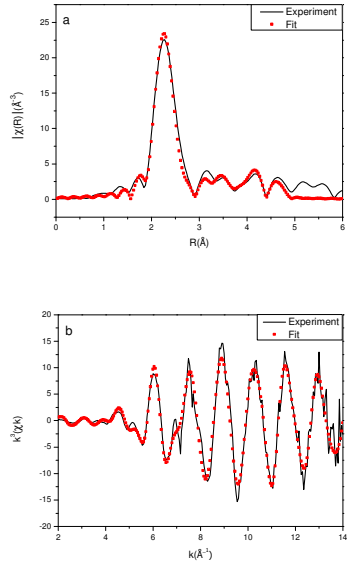

detected. For the sample with $x = 0.05$ annealed at 500 $^o$C small clusters and those with dimensions up to 15 nm are seen, and from EXAFS analysis almost 50 % of ZB and hexagonal inclusion have been obtained.

Table 4. The percentage of Mn atoms in ZB and hexagonal structure as estimated by EXAFS analysis.

|  | Annealed at 500 $^o$C | | Annealed at 600 $^o$C | |
| --- | --- | --- | --- | --- |
| Sample | ZB | Hexagonal | ZB | Hexagonal |
| $x = 0.025$ | 100% | 0 | 100% | 0% |
| $x = 0.04$ | 72(2)% | 28% | 40(5)% | 60% |
| $x = 0.05$ | 53(5)% | 47% | 15(5)% | 85% |
| $x = 0.063$ | 69(2)% | 31% | 50(5)% | 50% |

Fig. 11. Ga$_{0.95}$Mn$_{0.05}$As annealed at 600 °C. Model of 15 % (Mn,Ga)As ZB and 85% MnAs hexagonal inclusions. (a) Fourier Transform of the EXAFS (squares) and the result of fitting (full line). (b) EXAFS oscillations (squares) and the result of fitting (full line).

EXAFS analysis of the samples with different Mn content and after different annealing procedures excluded the existence of MnAs ZB nanoclusters in GaAs matrix. In all of the investigated samples the ZB (Mn,Ga)As nanoclusters were found. Due to small dimension of detected ZB nanoclusters (up to 6 nm) the Ga atoms in second sphere were detected in all samples. The ratio of unit cells exposed to the surface to those in bulk is about 0.5 for clusters up to 5 nm; therefore half of the Mn atoms would see Ga from the surface. However, if MnAs ZB is formed inside the cluster it would be visible in EXAFS as well. Nevertheless, when the dimensions of clusters reach 10 nm and the influence of the surface is negligible, hexagonal inclusions are formed. In the sample with the smallest Mn concentration ($x = 0.025$) annealed at 500 and 600 $^o$C, only ZB phase inclusions were found in agreement with TEM observation.

Examining the parameters of the fitting procedures collected in Tables 2 and 3, one may notice that for ZB clusters the disorder parameter $\sigma^2$ is rather high for the second and third sphere. This is obviously related to the small dimension of these clusters and distribution of the Mn content. In the case of hexagonal nanoclusters the relatively high disorder was found in the forth sphere composed of As atoms with the distance close to As atoms in ZB clusters.

Let's analyze the data collected in Table 4 showing the fraction of Mn atoms in ZB and hexagonal nanoclusters together with the dimensions of clusters presented in Fig. 7. One can see that for samples with small amounts of Mn, $x = 0.025$ small clusters with ZB structure were

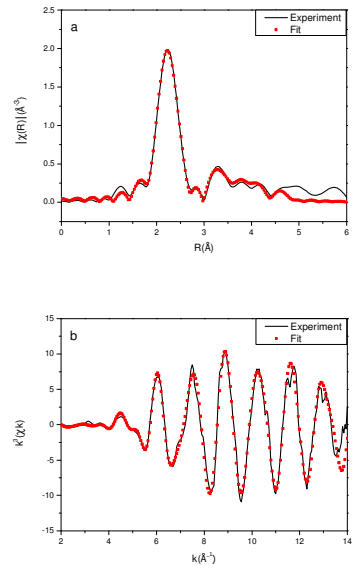

Fig. 12. Ga$_{0.937}$Mn$_{0.063}$As annealed at 500 °C. Model of 69 % (Mn,Ga)As ZB and 31% MnAs hexagonal inclusions. (a) Fourier Transform of the EXAFS (squares) and the result of fitting (full line). (b) EXAFS oscillations (squares) and the result of fitting (full line).

After annealing at 600 $^o$C most clusters have dimensions between 10 and 20 nm and EXAFS found only 15% of ZB clusters. In the case of the sample with $x = 0.063$ after annealing at 500 $^o$C the histogram shows that most clusters have dimensions between 5 and 10 nm, and EXAFS found 69% of atoms in ZB clusters. For the sample annealed at 600 $^o$C still almost 50% small clusters were seen, and EXAFS detected 50% of Mn atoms in ZB clusters. Therefore, our studies confirmed what was reported in many papers that small clusters (here up to 6-8 nm) in GaAs matrix have ZB structure and larger ones hexagonal structure. Moreover, the annealing at 600$^o$C is not sufficient to convert all clusters into hexagonal inclusions.

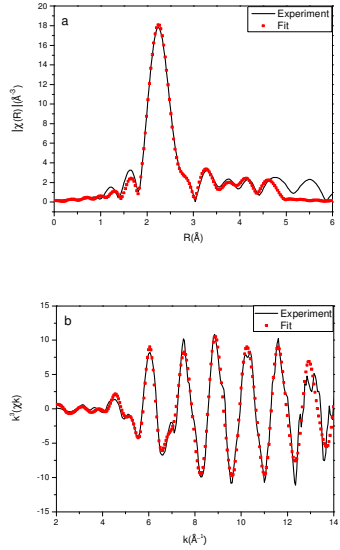

Fig. 13. Ga$_{0.937}$Mn$_{0.063}$As annealed at 600 °C. Model of 50 % (Mn,Ga)As ZB and 50% MnAs hexagonal inclusions. (a) Fourier Transform of the EXAFS (squares) and the result of fitting (full line). (b) EXAFS oscillations (squares) and the result of fitting (full line).

In the ref. 35 the EXAFS studies as-grown samples from the same series are presented and the fractions of Mn in the substitutional and interstitial position are estimated. Only in the sample $x = 0.025$, 90% of Mn atoms were located in the Ga position. In all other samples more than 40% of Mn atoms were in the interstitial position. After annealing of the sample with $x = 0.025$, only ZB inclusions were formed. In the case of other samples, annealed at 500 °C some hexagonal inclusions were formed. It may imply that Mn in Ga position is more difficult to convert into hexagonal inclusions even by annealing at 600 °C, but Mn in interstitial position easily forms hexagonal inclusions already during 500 °C annealing. The Mn-As bond length in the substitutional position in as-grown samples was found to be 2.48 Å. The Mn-As bond in the ZB inclusions is longer ~2.55 Å and in the case of sample with only ZB inclusions still the ferromagnetic properties at room temperature were observed.

### 3.5 Magnetic studies

In order to get more information about inclusions properties the structural study was complemented by magnetization measurements. As it was mentioned in the introduction both hexagonal MnAs and cubic GaMnAs are ferromagnets, however critical temperature of MnAs is about 320 K, while for the GaMnAs it varies with Mn concentration and hardly reaches 190 K [15]. Therefore above 200 K GaMnAs behaves as a paramagnet, while hexagonal MnAs is still in a ferromagnetic state.

The magnetic behavior of our composite in two extreme cases is illustrated in Fig. 14 and 15, where magnetization of the samples with $x = 0.05$ (Fig. 14) and, $x = 0.025$ (Fig.15) annealed at 600 °C is plotted as a function of magnetic field for several temperatures.

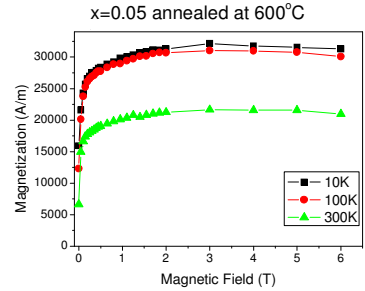

Fig. 14. Magnetization of Ga$_{0.95}$Mn$_{0.05}$As annealed at 600 °C as a function of magnetic field measured at 10 K, 100K and 300K.

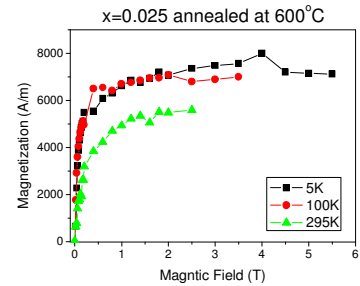

Fig. 15. Magnetization of Ga$_{0.975}$Mn$_{0.025}$As annealed at 600 °C as a function of magnetic field measured at 5 K, 100K and 295 K.

In the case of the sample containing nearly solely hexagonal MnAs inclusions discussed above ($x = 0.05$) magnetization shows typical ferromagnetic-like behavior, (i.e. very fast saturation with magnetic field) even at 300 K, in agreement with magnetic properties of bulk MnAs. Similar fast saturation of magnetization with magnetic field is also observed for the sample containing solely cubic (Mn,Ga)As inclusions (Fig. 15). While it is not surprising at low temperatures, where inclusions built of ferromagnetic GaMnAs may reveal such behavior, very fast magnetizing of the system at 300 K, where bulk GaMnAs is paramagnetic, is not expected. The observed result can be interpreted assuming that Mn ions inside the inclusions are still ferromagnetically coupled, even at room temperature. It should be stressed that inclusions in this sample were proved by EXAFS experiment (see discussion in Sec. 3.4) to be built of (Mn,Ga)As, not of MnAs. This may suggest that inclusions in this sample are formed of GaMnAs with Mn concentrations higher than 10% (highest in the GaMnAs epilayers), yet still well below 100% (MnAs). Theoretical calculations performed by K. Sato [42] showed that sample containing solely cubic (Ga,Mn)As with content of Mn higher than 15% are still ferromagnetically coupled, even at room temperature.

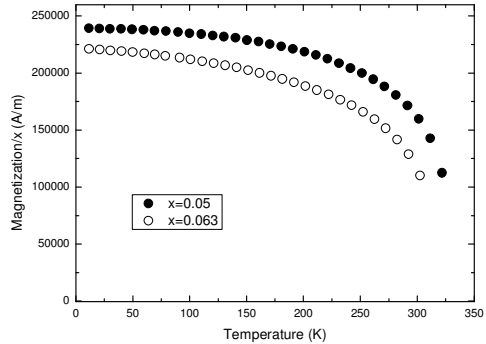

Fig. 16. Magnetization per Mn content versus temperature shown for samples with different content of hexagonal MnAs inclusions.

Typical magnetic behavior of our composite containing different amount of hexagonal inclusions is illustrated in Fig.16, where magnetization scaled with Mn concentration (i.e. M/$x$) for the samples with $x = 0.05$ and $x = 0.063$ annealed at 600 $^{o}$C, is plotted as a function of temperature. The presented data were collected after cooling the samples at low magnetic field to minimize the anisotropy effects, typical for a superparamagnet. The details of the experiment will be published elsewhere. The sample containing nearly solely hexagonal MnAs inclusions ($x = 0.05$) reveals typical ferromagnetic behavior, with critical temperature not less than 320 K (upper limit of the magnetometer used). For the sample with $x = 0.063$ containing 50% of hexagonal MnAs inclusions and 50% of cubic (Mn,Ga)As the situation is essentially the same, except that measured magnetization is smaller by 10-15%. Such behavior should be expected for the system containing two magnetic phases. Strong ferromagnetic contribution of hexagonal MnAs overrides the contribution of (Mn,Ga)As, however relative concentration of hexagonal MnAs inclusions is smaller in this sample, which results in smaller total magnetization. In other words observed smaller absolute values of M/$x$ reflect smaller abundance of hexagonal MnAs phase in this sample.

## 4. Conclusions

The set of Ga$_{1-x}$Mn$_x$As layers grown on GaAs (001) substrate with different contents of Mn and annealed at 500 °C or 600 °C was studied. For samples annealed at lower temperatures, creation of mainly cubic nanoinclusions was observed. In the case of layers with small Mn concentration annealed at higher temperatures, formation of cubic precipitations was still preferable. However, for the samples with higher Mn concentration, already at 500 $^{o}$C hexagonal inclusions were formed, and annealing at 600° C led to creation of a higher number of hexagonal nanoinclusions. The strain of the GaAs matrix related to the creation of Mn-containing cubic nanoclusters was larger than for hexagonal ones, which indicates that cubic inclusions act with larger stress on the matrix unit cells than the hexagonal nanoinclusions.

XAFS studies rule out the possibility of formation of a hypothetic MnAs ZB compound, widely discussed in the literature. Only (Mn,Ga)As ZB clusters were detected. In all samples, in order to get reasonable fit of EXAFS data to the model, the Ga atoms have to be located in the second coordination sphere around Mn atoms instead of Mn as in MnAs ZB. The comparison of the distribution of nanocluster dimensions estimated from TEM with the fraction of Mn atoms found by EXAFS in ZB and hexagonal clusters confirmed that clusters up to 6-8 nm have ZB structure and larger ones hexagonal structure. The fraction of Mn found in hexagonal structure was directly related to magnetic properties of samples. Moreover, the ZB inclusions found in the sample containing solely cubic (Mn,Ga)As inclusions remain ferromagnetic even at room temperature.

Based on X-ray, EXAFS, and TEM results the lattice strain introduced to the GaAs matrix related to creation of both kind of nanoinclusion was estimated.

**Acknowledgements:** This work was partially supported by national grant of Ministry of Science and High Education N202-052-32/1189 as well as by DESY/HASYLAB (EC support program: Transnational Access to Research Infrastructures) and the European Community under Contract RII3-CT-2004-506008 (IA-SFS).